\documentclass[preprint,12pt]{elsarticle}

\usepackage{graphicx}

\usepackage{subcaption}

\usepackage{hyperref}
\hypersetup{pdfauthor={Guo-Xiang Zhi, Chen-Chao Xu, Si-Qi Wu, Fanlong Ning, Chao Cao}}

\usepackage{amssymb}
\usepackage{amsmath}
\usepackage[mathlines,displaymath]{lineno}

\usepackage{xcolor}
\definecolor{red}{rgb}{1,0,0}
\definecolor{blue}{rgb}{0,0,1}
\definecolor{black}{rgb}{0,0,0}
\usepackage{soul}

\usepackage{booktabs}

\newcommand\blfootnote[1]{%
  \begingroup
  \renewcommand\thefootnote{}\footnote{#1}%
  \addtocounter{footnote}{-1}%
  \endgroup
}

\newcounter{bla}

\journal{Computer Physics Communications}

\begin{document}

\begin{frontmatter}

\title{WannSymm: A symmetry analysis code for Wannier orbitals}

\author[a]{Guo-Xiang Zhi\fnref{1}}
\author[a]{Chenchao Xu\fnref{1}}
\author[a]{Si-Qi Wu}
\author[a]{Fanlong Ning}
\author[a,b]{Chao Cao\corref{author}}

\address[a]{Department of Physics and Center for Correlated Matter, Zhejiang University, Hangzhou 310013, P. R. China}
\address[b]{Condensed Matter Group, Department of Physics, Hangzhou Normal University, Hangzhou 311121, P. R. China}
\fntext[1]{These authors contribute equally in this work.}
\cortext[author] {Corresponding author. \\ \textit{E-mail address:} ccao@zju.edu.cn}

\begin{abstract}
    \blfootnote{\copyright\ 2021 This manuscript version is made available under the CC-BY-NC-ND 4.0 license https://creativecommons.org/licenses/by-nc-nd/4.0/}
    We derived explicit expressions of symmetry operators on Wannier basis, and implemented these operators in WannSymm software. Based on this implementation, WannSymm can i) symmetrize the real-space Hamiltonian output from Wannier90 code, ii) generate symmetry operators of the little group at a specific k-point, and iii) perform symmetry analysis for Wannier band structure. In general, symmetrized Hamiltonians yield improved results compared with the original ones when they are employed for nodal structure searching, surface Green's function calculations, and other model calculations. 
\end{abstract}


\end{frontmatter}

{\bf PROGRAM SUMMARY/NEW VERSION PROGRAM SUMMARY}

\begin{small}
\noindent
{\em Program Title:}  WannSymm \\
{\em Licensing provisions(please choose one): }    BSD 3-clause \\
{\em Programming language:  } C \\
{\em External Libraries:  } spglib; BLAS and LAPACK; MPI libraries (optional) \\
{\em RAM: } Depend on the number of orbitals, number of real-space lattice points and order of spatial group. \\
{\em Nature of problem(approx. 50-250 words): } Generate the symmetry operators in Wannier basis; calculate the symmetry eigenvalues and characters; symmetrize the real-space Hamiltonian according to the crystal structure. It deals with nonmagnetic or magnetic systems with or without spin-orbit coupling. For magnetic systems, it can deal with ferromagnetic, commensurate antiferromagnetic long range ordered collinear or non-collinear systems with spin-orbit coupling.\\
{\em Running time:} All of the examples run in less than 3 minutes on a two-way Intel Xeon 2650v4 node (24 core). \\

\end{small}


\section{Introduction}
\label{SecIntroduction}
Symmetry serves as one of the most fundamental concept and properties in the modern physics. In the condensed matter physics, in particular, the Landau's symmetry breaking theory\cite{landau1937theory1,landau1937theory2} for phase transition is widely applied in the second order phase transitions such as superconductivity phase transitions and magnetic phase transitions, where the symmetry-based order parameter signals the phase transition. In the solid state theory, the irreducible representations of a space group can be used in labelling the electronic energy bands in a crystalline solid\cite{GroupTheoryBradley}, which becomes particularly important in the modern band topology theory. The discovery of topological insulators\cite{TI-review-Hasan,TI-review-Fu,TI-SnTe}, quantum anomalous Hall insulators\cite{TI-AHE}, as well as topological Weyl and Dirac semimetals brings intensive study on the topological properties of materials\cite{RevWeylDiracSemimentals}, where the band symmetry is crucial to the determination of band topologies. Such topology-symmetry correspondence became evident in the more recently discovered symmetry based indicator theory that can be used to distinguish topological non-trivial phases from trivial ones\cite{TopoClassEigenSlager,TopoIndi-Z2,TopoIndi-Watanabe230,TopoIndi-inv4SC}. This is particularly useful for superconductors, because one may apply the symmetry-indicator theory and predict topological superconductor candidates using normal state Hamiltonian combined with its superconducting pairing symmetry, if weak-coupling condition is met\cite{TopoIndi-inv4SC}. Furthermore, for most ab initio codes, symmetry is used to improve calculation accuracy, as well as to reduce the calculation burden. 

Many state-of-art first-principles calculation codes utilizes plane-wave or augmented plane-wave basis, which are more expensive than localized basis. Therefore in many cases, a real-space Hamiltonian is obtained via Wannier projection (e.g., Wannier90 code\cite{Code-Wannier90}) after the ground state is converged, and the post-process calculations employs this real-space Wannier Hamiltonian to speed up the calculations. However, due to numerical errors introduced in the projection procedure, the crystal symmetries may be violated in the derived real-space Hamiltonian. These numerical errors may lead to incorrect results. Therefore, it is important to restore correct symmetries in the derived real-space Hamiltonian. At present, there are several codes performing similar tasks.  For example, the \textsc{wannhr\_symm} code in the \textsc{WannierTools} package\cite{Code-WannierTools} performs symmetrization of Hamiltonian without magnetic orders; the symmetry-adapted Wannier function method\cite{SymmAdaptedWannFun} does wannierization with symmetry constraint for spinless Hamiltonians.

In this article, we introduce the WannSymm code which can be applied to non-magnetic (NM), ferromagnetic (FM) or anti-ferromagnetic (AFM) crystals. It can symmetrize the real-space Wannier Hamiltonian,  perform symmetry analysis and obtain the eigenvalues and characters of relevant symmetries at a given $\textbf{k}$ point. It implements parallel algorithm for symmetrizing the real-space Hamiltonian. The method does not depend on the radial part of Wannier orbitals, and thus it can be applied to any tight-binding like Hamiltonian defined on a local basis-set with atomic-like angular dependence. Using WannSymm, we have already performed symmetry analysis for the unconventional superconductor K$_{2}$Cr$_{3}$As$_{3}$ \cite{Xu-233-cal-topo}, calculated the simple $\mathcal{Z}_{2}$ index for topological material W$_{2}$As$_{3}$ \cite{W2As3}, the mirror eigenvalues for nodal ring semimetal InTaSe$_{2}$\cite{LI2021243} and correctly identify the nodal structures and topological invariants for the Kondo nodal-line semimetal Ce$_{3}$Pd$_{3}$Bi$_{4}$\cite{Ce343prl}.

\section{Algorithm background}
\label{SecAlgo}

\subsection{Crystal symmetry operations}
The WannSymm code employs spglib\cite{Lib-spglib} to determine the relevant symmetry operations for a specific system. The spglib obtains the symmetry operations from the input of crystal structures. However, these symmetry operations are purely spatial and do not involve time-reversal operation, forming a group denoted as $\mathcal{G}_S$. For the systems without long-range magnetic order, the full symmetry of the Hamiltonian is then $\mathcal{G}_H=\mathcal{G}^T=\mathcal{G}_S\oplus\mathcal{TG}_S$, where $\mathcal{T}$ is the time-reversal symmetry. For the systems with long-range magnetic order [either ferromagnetic (FM) or antiferromagnetic (AFM)], the rotation operations are required to act on both spatial space and spin space when spin-orbit coupling (SOC) is considered. Therefore, the full symmetry of the Hamiltonian $\mathcal{G}_H$ is a subgroup of $\mathcal{G}^T$. Noticing that the operations in $\mathcal{G}_H$ should leave the magnetic moment unchanged as well, we determine the elements of $\mathcal{G}_H$ by taking every $\mathcal{R}\in\mathcal{G}^T$ and operating it on the magnetic sublattice. If the sublattice after the operation $\mathcal{R}$ remains equivalent to the original one,  $\mathcal{R}$ belongs to $\mathcal{G}_H$. In WannSymm, the magnetic sublattice is represented by a set of magnetic moment vectors at the atomic positions $\mathbf{\tau}$.

\subsection{Atomic-like Wannier basis and Real-space Hamiltonian}
\label{SecWannierBasis}
In practice, Wannier orbitals are usually obtained by projecting Bl\"{o}ch waves to atomic orbital-like initial guess $|g\rangle$. It is our experience that the angular dependence of $|g\rangle$ is more important than its radial function in order to obtain an accurate Wannier fitting in most cases. In addition, we shall see that the WannSymm code relies on the angular dependence of the Wannier basis to perform the symmetrization procedure. Therefore, we shall assume that the Wannier basis are cubic-harmonic atomic-orbital like, i.e. 

\[| w_{i {\bf R}} \rangle = | {\bf R}  \tau n l m \sigma\rangle\]
where $i$ denotes the $i$th orbital,  $\bf R$ is Bravais lattice vector, ${\bf \tau}$ is the atomic position of the $i$-th orbital in the unit cell, and $n, l$ and $\sigma$ are its primary, orbital and spin quantum numbers, respectively.  $m$ is the index of the cubic harmonic. The corresponding real-space Hamiltonian can 
be written as
\[ H^{\bf R}_{ij}=\langle w_{i {\bf 0}}|\hat{H}| w_{j {\bf R}}\rangle = \langle {\bf 0 \tau}_i n_i l_i m_i \sigma_i| \hat{H}|{\bf R \tau}_j n_j l_j m_j \sigma_j \rangle \]

Following the definition of Fourier transformation, we have the basis of corresponding reciprocal-space and Hamiltonian in this basis

\begin{equation}
\label{basis_R_Fourier}
|w_{i {\bf k}}\rangle=\sum_{{\bf R}} e^{\mathrm{i} {\bf k} \cdot {\bf R}} |w_{i {\bf R}}\rangle
\end{equation}
\begin{equation}
\label{ham_R_Fourier}
H^{{\bf k}}_{ij} = \sum_{{\bf R}} e^{\mathrm{i} {\bf k}\cdot{\bf R}} H^{{\bf R}}_{ij} 
\end{equation}

We use the same convention as employed in Wannier90\cite{Code-Wannier90} code. If the tight-binding convention instead of Wannier convention is employed, an additional $e^{\mathrm{i}\mathbf{k}\cdot\mathbf{\tau}_i}$ or $e^{\mathrm{i} \mathbf{k}\cdot(\mathbf{\tau}_j-\mathbf{\tau}_i)}$ factor may appear in the basis or Hamiltonian transformation, respectively. 

\subsection{Symmetry operator}

The matrix representation of symmetry operators are derived in the Hilbert space formed by the atomic-like Wannier basis $|\mathbf{R \tau} n l m \sigma \rangle$. The rotation operator $\hat{D}(\mathcal{R})$ corresponding to a proper rotation $\mathcal{R}$ is

\[\hat{D}(\mathcal{R}) = \exp \left(-i \phi \frac{\mathbf{\hat n} \cdot \hat{\mathbf L }}{\hbar} \right)\]
where $\mathbf{\hat n}$ is the normalized rotation axis, $\phi$ is the rotated angle, and $\hat{\mathbf{L}}$ is the angular momentum operator. For spherical harmonics, it is straightforward to obtain the matrix form of $\hat{\mathbf{L}}$, hence the rotation matrix $\mathcal{D}^l$ in spherical harmonics. Therefore we employ the transformation $U$ from cubic harmonics to spherical harmonics, defined by 
\[|Y(l, m)\rangle=\sum_{m'}U_l^{mm'}|lm'\rangle\]
where $|Y(l, m)\rangle$ and $|lm'\rangle$ denote the spherical harmonics and cubic harmonics, respectively. The indices of lattice, position, primary quantum number and spin are omitted in the expression. The rotation matrix in cubic harmonics $D^l$ is then given by $D^l=U_l\mathcal{D}^lU_l^{-1}$. 

We denote the rotation matrix of a specific rotation $\mathcal{R}$ and angular quantum number $l$ as $D^l(\mathcal{R})$, and the spinor part as $D^s(\mathcal{R})$. 
Therefore, 

\[\hat{\mathcal{R}}|\mathbf{R\tau} nlm\sigma\rangle=\sum_{m'\sigma'} D^l(\mathcal{R})_{m'm}D^s(\mathcal{R})_{\sigma'\sigma}|\mathbf{R'\tau'} nlm'\sigma'\rangle\]
where $\mathbf{R'+\tau'}\equiv \mathcal{R}(\mathbf{R+\tau})$ is the new atomic position after rotation.

In addition, the spatial inversion operator simply introduces a factor of $(-1)^l$ for the orbital part. For the time-reversal operation $\mathcal{T}=\mathcal{U\cdot K}$,  since the cubic harmonics are real functions, it has trivial effect on the orbital part during the symmetrization.  Therefore, we only need to take care of the spinor part using $-i\sigma_y\mathcal{K}$ with $\mathcal{K}$ being the complex conjugation. With these definitions, the matrix representation of all symmetry operations can be fully constructed. 

\subsubsection{Real space}

We then elaborate a little bit further of the symmetry operator for the real-space Hamiltonian, which is useful during the symmetrization process. According to above derivation, a symmetry operation $\mathcal{S}$ acting on the Wannier basis in general can be written as:
\[\hat{\mathcal{S}}|\mathbf{R\tau} nlm\sigma\rangle=\sum_{m'\sigma'} S^l_{m'\sigma' m\sigma} |\mathbf{R'\tau'} nlm'\sigma'\rangle\]
Thus, $\langle \mathbf{R'\tau'} n'l'm'\sigma' | \hat{\mathcal{S}} | \mathbf{R\tau} nlm\sigma\rangle=S^l_{m'\sigma' m\sigma} \delta_{nn'}\delta_{ll'}\delta_{\mathcal{S}(\mathbf{R+\tau}),(\mathbf{R'+\tau'})}$. Therefore, the real-space symmetry operation matrices are very sparse.

Since the Hamiltonian $\hat{H}$ transforms as $\hat{\mathcal{S}} \hat{H} \hat{\mathcal{S}}^{\dagger}$ and the elements of real-space Hamiltonian are $H_{ij}^{\mathbf{R}}=\langle\mathbf{0 \tau}_i n_i l_i m_i \sigma_i | H | \mathbf{R \tau}_j n_j l_j m_j \sigma_j \rangle$, the rotated Hamiltonian is then

\[ \left\lbrace \mathcal{S}H\mathcal{S}^{\dagger}\right\rbrace_{ij}  =  \langle \mathbf{0\tau}_i n_i l_i m_i \sigma_i | \mathcal{S}H\mathcal{S}^{\dagger} | \mathbf{R \tau}_j n_j l_j m_j \sigma_j \rangle \]
\[ = \sum_{i'j'}S^{l_i}_{m_i\sigma_i m_{i'} \sigma_{i'}}\delta_{\mathcal{S}(\mathbf{R'+\tau}_{i'}),\mathbf{\tau}_i} \delta_{n_i n_{i'}} \delta_{l_i l_{i'}}\langle \mathbf{R' \tau}_{i'} n_{i'} l_{i'} m_{i'} \sigma_{i'} | H |  \mathbf{R'' \tau}_{j'} n_{j'} l_{j'} m_{j'} \sigma_{j'} \rangle \times\]
\[(S^{l_j\dagger})_{m_{j'} \sigma_{j'} m_j\sigma_j} \delta_{\mathcal{S}(\mathbf{R''+\tau}_{j'}),\mathbf{R+\tau}_j} \delta_{n_j n_{j'}} \delta_{l_j l_{j'}}=\sum_{i'j'} S^{l_i}_{m_i\sigma_i m_{i'} \sigma_{i'}} H_{i'j'}^{\mathbf{R''-R'}} (S^{l_j\dagger})_{m_{j'} \sigma_{j'} m_j\sigma_j} \times\]
\[\delta_{\mathbf{R'+\tau}_{i'},\mathcal{S}^{-1}\mathbf{\tau}_i} \delta_{\mathbf{R''+\tau}_{j'},\mathcal{S}^{-1}\mathbf{R}+\mathcal{S}^{-1}\tau_j} \delta_{n_i n_{i'}} \delta_{l_i l_{i'}} \delta_{n_j n_{j'}} \delta_{l_j l_{j'}} \]

In general, a symmetry operation $\mathcal{S}$ maps an atom $i$ at $\tau_i$ to its symmetrically equivalent site with an optional lattice translation $\mathbf{R}_i^{\mathcal{S}}$, which depends only on the symmetry operation $\mathcal{S}$ and the site coordinate $\tau_i$. With this in mind, we have $\mathbf{R''-R'}=\mathcal{S}^{-1}\mathbf{R}+\mathbf{R}_j^{\mathcal{S}^{-1}}-\mathbf{R}_i^{\mathcal{S}^{-1}}$. 

Using the above result, we are able to calculate the real-space Hamiltonian after symmetry operation $\mathcal{S}$. This procedure is employed in the Hamiltonian symmetrization process, which involves:

\[H^{\mathrm{sym}}=  \frac{1}{|\mathcal{G'}_H|} \sum_{\mathcal{S}\in\mathcal{G'}_H}  \mathcal{S}H\mathcal{S}^{\dagger}\]
where $\mathcal{G}_H'=\mathcal{G}_H/\mathcal{G}_T$ with $\mathcal{G}_T$ being all the integer tanslations. This is because all the integer translational symmetries are automatically preserved in the Wannier representation.

\subsubsection{Reciprocal space}

According to Eq. \ref{basis_R_Fourier}, the rotation of reciprocal Wannier orbitals are

\[\hat{\mathcal{S}}| w_{i\mathbf{k}} \rangle= \sum_{\mathbf{R}} e^{\mathrm{i} \mathcal{S}\mathbf{k}\cdot (\mathcal{S}\mathbf{R})} \hat{\mathcal{S}} | \mathbf{R} \tau nlm\sigma \rangle = \sum_{\mathbf{R'}} e^{\mathrm{i} \mathbf{k'}\cdot\mathbf{R'}} \sum_{m'\sigma'} S^l_{m'\sigma' m\sigma} | \mathbf{R''}\tau' n l m'\sigma'\rangle \]
where $\mathbf{k'}=\mathcal{S}\mathbf{k}$, $\mathbf{R'}=\mathcal{S}\mathbf{R}$ and $\mathbf{R''}+\tau'=\mathcal{S}(\mathbf{R}+\tau)=\mathbf{R'}+\mathbf{R}^{\mathcal{S}}_{\tau}+\tau'$. As a result,

\begin{align}
    \label{Sk_expression}
    \hat{\mathcal{S}} | w_{i\mathbf{k}}\rangle &=\sum_{m'\sigma'} S^l_{m'\sigma' m\sigma} \sum_{\mathbf{R'}} e^{\mathrm{i} \mathbf{k'}\cdot\mathbf{R'}} | \mathbf{R'}+\mathbf{R}^{\mathcal{S}}_{\tau} \tau' n l m' \sigma'\rangle \nonumber \\
    &=\sum_{m'\sigma'} S^l_{m'\sigma' m\sigma} e^{-\mathrm{i}\mathbf{k'}\cdot\mathbf{R}^{\mathcal{S}}_{\tau}} |w_{i'\mathbf{k'}}\rangle 
\end{align}

Thus, using $S^{\mathbf{k}}_{ij}=\langle w_{i\mathbf{k}} |\hat{\mathcal{S}}| w_{j\mathbf{k}}\rangle$, we can calculate $H^{\mathbf{k'}}$ ($\mathbf{k'}=\mathcal{S}\mathbf{k}$) from $H^{\mathbf{k}}$ using $H^{\mathbf{k'}}=S^{\mathbf{k}} H^{\mathbf{k}} (S^{\mathbf{k}})^{\dagger}$. 

In addition, $S^{\mathbf{k}}$ can be used to perform symmetry analysis of a Bl\"{o}ch state. In order to do that, we first identify the little group $\mathcal{G}^{\mathbf{k}}_H$ of Hamiltonian $H_{\mathbf{k}}$ at a specific point $\mathbf{k}$. The eigenstates of $H_{\mathbf{k}}$ therefore can block-diagonalize $S^{\mathbf{k}}$, with the trace of each block being the characters of different symmetry representations. Thus the irreducible representation of each Bl\"{o}ch state can be determined using a character table.

\subsection{Details of the code}
\subsubsection{Data Structure}
In the WannSymm code, each symmetry operation is represented by two flags of spatial inversion and time-reversal in addition to a proper rotation and the associated translation. 

In order to implement the above algorithm, the real-space Hamiltonian $H^{\mathbf{R}}_{ij}=\langle w_{i\mathbf{0}} | H | w_{j\mathbf{R}}\rangle$ are divided into blocks such that the bra/ket states in each block have the same $\tau$, $n$, and $l$. In other words, the Hilbert space is divided into subspaces formed by atomic orbitals with the same primary and orbital quantum numbers on each atom. A mapping between subspace orbital indices to global orbital indices is maintained by the WannSymm code to facilitate the calculation. 

For the reciprocal space calculations, since the $S^{\mathbf{k}}$ has a definite dimension the same as the $H^{\mathbf{k}}$, we can construct $S^{\mathbf{k}}$ using Eq. \ref{Sk_expression} with the assist of the subspace-global mapping.

\begin{figure}
    \centering
    \includegraphics[width=1.0\textwidth]{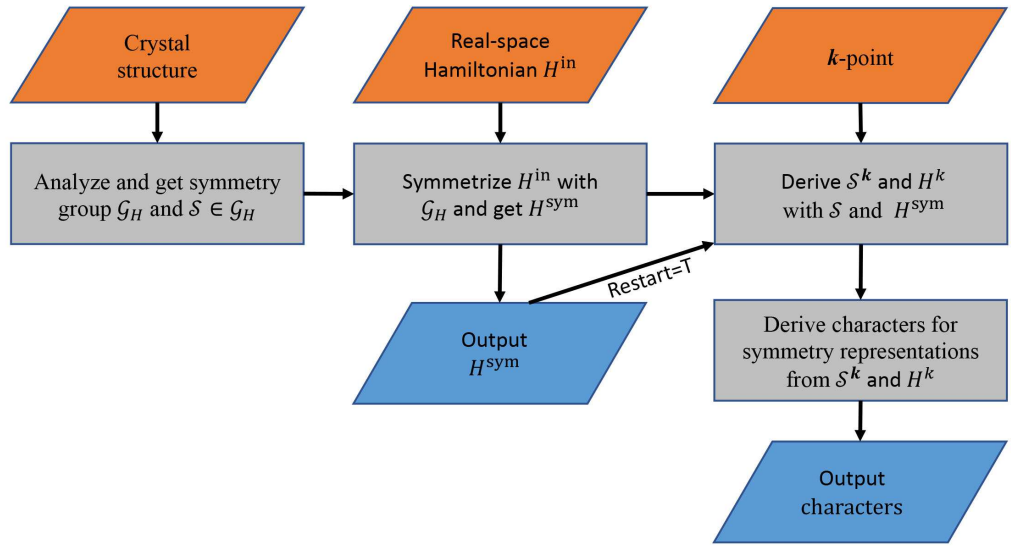}              
    \caption{
      Simplified schematic chart of the program. Note that the calculation of characters for symmetry representations of a given $\mathbf{k}$-point is optional.
    }
    \label{flowchart}
\end{figure}

The structure of the program is illustrated in Fig. \ref{flowchart}.

\subsubsection{Installation and Typical Input/Output File}
The WannSymm code depends on spglib to analyze the symmetry of crystal. In addition, it also requires a working implementation of LAPACK/BLAS library. Makefile compiling system is used. In order to compile the WannSymm code, the users need to change the make.sys file in the top folder, and provide paths of spglib and LAPACK/BLAS library. 

The WannSymm code can also utilize message passing interface (MPI) to speed up the calculations. Currently, MPI parallelization is implemented over the number of symmetry operations. 

The compiled code can be executed with 

        {\tt mpirun -np \${Num\_of\_process} wannsymm.x \${InputFile} }
        
where {\tt \${InputFile}} is the name of the master input file. If it is not provided, a default master input file {\it wannsymm.in} must be provided.  A typical master input file looks like:

\begin{verbatim}
# template input file of wannsymm.x
# anything following '#', '!' or '//' in a line will be regard as comments
# tag names are case insensitive( SeedName and seednAme are equivalent)
# 

DFTcode  = VASP

Spinors  = T

SeedName ='MnF2'

# Input file for crystal structure
Use_POSCAR = 'POSCAR'

# Projections_in_Format_of_wannier90
# NOTE: Do not support local axis now
begin projections
  Mn: s;d
   F: p
end projections

#Use_Symmetry = 'symmetries.in'

# Magnetization specification
MAGMOM = 0 0 5 0 0 -5 12*0
#SAXIS  = 0 0 1
#symm_magnetic_tolerance = 1E-4

#restart = T
# Kpoint for calculating characters
#kpt = 0 0 0

\end{verbatim}

The master input file must contain {\tt DFTcode} tag, {\tt SeedName} tag, {\tt Spinors} tag, projections block and crystal structure information. Most of these tags are self-explanatory, and we shall not restate their meanings here. The DFTcode tag is related to the type of the real-space Hamiltonian from Wannier90 code. It can be either "VASP" for \emph{ Vienna Abinitio Simulation Package} (VASP)\cite{Code-VASP1993,Code-VASP1999} users or "QE" for Quantum ESPRESSO\cite{Code-QE} users. For the Wien2K \cite{Code_wien2k} users, the type of the real-space Hamiltonian can be specified as the VASP-type. The projections block follows the Wannier90 convention, and the crystal structure information can be provided either in the POSCAR-format file specified in the {\tt Use\_POSCAR} tag or in the {\tt Structure\_in\_Format\_of\_POSCAR} tag. In addition, a real-space Hamiltonian following the Wannier90 convention should be provided in {\it seedname\_hr.dat} file as well.
 
The execution generates several output files. The most important output file is {\it seedname\_symmed\_hr.dat}, which contains the symmetrized Hamiltonian. In addition, an output file {\it symmetries.dat} contains all the symmetries found by spglib. The output file {\it wannsymm.out} contains additional important details of the calculation. Finally, one can track the progress of the calculations by monitoring the {\it .progress-of-threadXX} files:

\begin{verbatim}
        tail -f .progress-of-thread1
\end{verbatim} 

After the symmetrized Hamiltonian is generated, the symmetry of Bl\"{o}ch states can be analyzed by setting additional two tags in the master input file:

\begin{verbatim}
Restart = T
Kpt = 0  0  0
\end{verbatim}

{\tt Restart} tag tells the code to enter analysis mode and skip the symmetrization procedure, whereas {\tt Kpt} tag specifies the K-point (in unit of reciprocal lattices) to be analyzed. Once the calculation is done, the code generates an additional output files {\it bnd\_sym\_characters}. It reads:

\begin{verbatim}
kpt:     0.000000000     0.000000000     0.000000000
Related symmetries:
symm   1: Identity
symm   2:  60.0 deg rot around ( 0.2500,-0.2588, 0.9330) with inv
symm   3: 120.0 deg rot around ( 0.2500,-0.2588, 0.9330)
symm   4: 180.0 deg rot around ( 0.2500,-0.2588, 0.9330) with inv
symm   5:-120.0 deg rot around ( 0.2500,-0.2588, 0.9330)
symm   6: -60.0 deg rot around ( 0.2500,-0.2588, 0.9330) with inv
symm   7: 180.0 deg rot around (-0.8700,-0.4830, 0.0991)
symm   8: 180.0 deg rot around (-0.9659, 0.0000, 0.2588) with inv
symm   9: 180.0 deg rot around (-0.8030, 0.4830, 0.3491)
symm  10: 180.0 deg rot around (-0.4250, 0.8365, 0.3459) with inv
symm  11: 180.0 deg rot around ( 0.0670, 0.9659, 0.2500)
symm  12: 180.0 deg rot around ( 0.5410, 0.8365, 0.0871) with inv
bnd    1:  2.000+0.000i,  1.732+0.000i,  1.000+0.000i, -0.000+0.000i,
           1.000-0.000i,  1.732-0.000i,  0.000+0.000i,  0.000+0.000i,
           0.000+0.000i,  0.000+0.000i,  0.000+0.000i,  0.000+0.000i,

bnd    2:  2.000+0.000i,  1.732+0.000i,  1.000+0.000i, -0.000+0.000i,
           1.000-0.000i,  1.732-0.000i,  0.000+0.000i,  0.000+0.000i,
           0.000+0.000i,  0.000+0.000i,  0.000+0.000i,  0.000+0.000i,
... ...
... ...
\end{verbatim}

The first line is the K-point calculated, followed by the little group information (all the symmetry operations of the little group). After that, the characters of each state are listed, which can be used with a character table to figure out the irreducible representation of the state.

\section{Examples and verification}
\label{SecExamples}

To verify the validity of our algorithm and code, we share here several examples in which our code is applied to symmetrize the Wannier Hamiltonian. These examples consists of four different materials, including paramagnet K$_2$Cr$_3$As$_3$, anti-ferromagnet MnF$_2$, ferromagnet CrO$_2$ and topological nodal-line semimetal Ce$_3$Pd$_3$Bi$_4$\cite{Ce343prl}.

 The real-space Hamiltonian employed in these examples are generated by the VASP, QE or Wien2K and their corresponding interface codes. In all these calculations, unless otherwise specified, the spin-orbit coupling (SOC) effect is considered as a second variation to the full Hamiltonian. All the initial guess of Wannier orbitals are atomic-like. During the Wannierization, we completely turn off the minimization of Wannier spreading, because the minimization procedure has no symmetry constraint, and may severely alter the symmetry (or the angular dependence of the orbitals). This procedure is conceptually the same as the maximally projected Wannier function proposed by Anisimov \emph{et~al.} \cite{MPRJ}. 
 
\subsection{Paramagnet \texorpdfstring{K$_2$Cr$_3$As$_3$}{K2Cr3As3} }
\label{SecEgK233}
K$_2$Cr$_3$As$_3$ received much attention and intensive study in recent years, due to its possible unconventional superconductivity and nontrivial topological property \cite{Xu-233-cal-topo,k233exp,k233cal}. Experimentally and theoretically, it was proposed to be paramagnetic at normal ground state. Structurally, it has quasi-one-dimensional structure with the space group $P\bar{6}m2$ (No. 187), which contains 12 symmetry operations. 

\begin{figure}[htbp]
  \begin{subfigure}{0.49\textwidth}
  	\includegraphics[width=\linewidth]{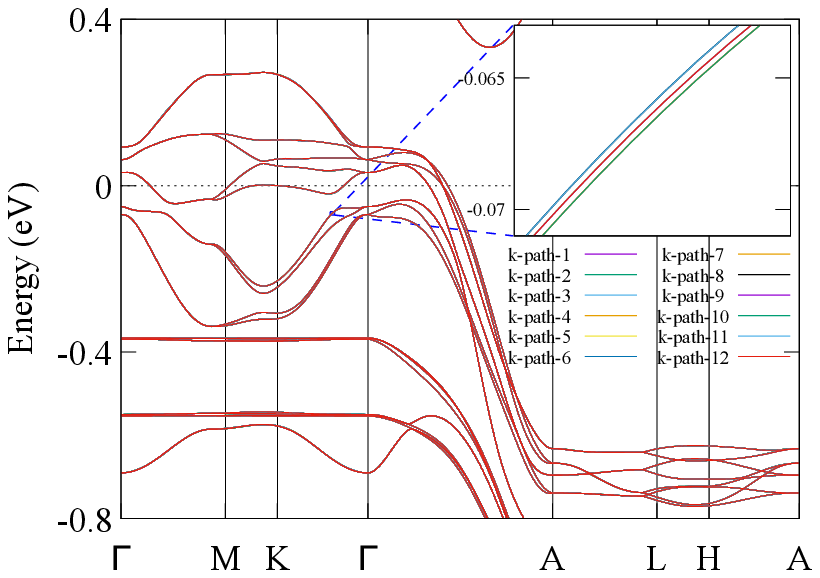}
 	  \caption{} 
  \end{subfigure}
  \hspace*{\fill} 
  \begin{subfigure}{0.49\textwidth}
  	\includegraphics[width=\linewidth]{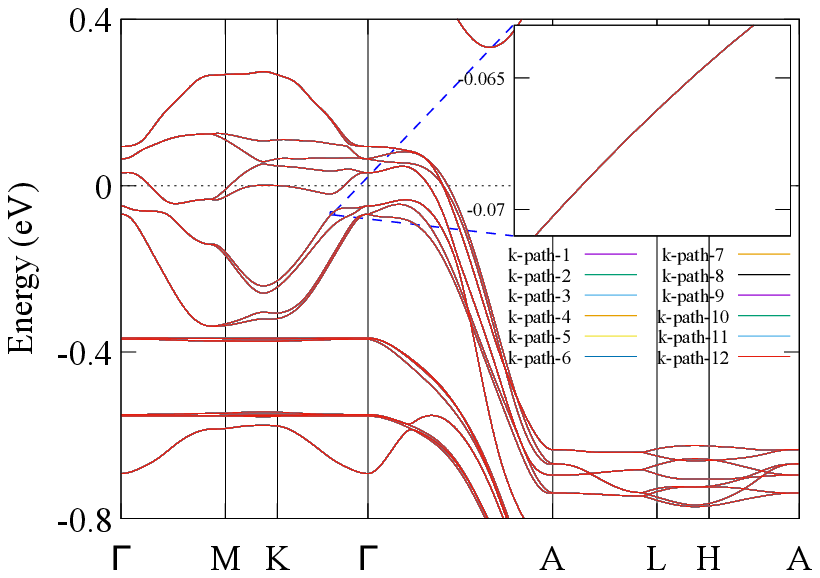}
 	  \caption{} 
  \end{subfigure}
  \caption{Band structures of K$_2$Cr$_3$As$_3$ calculated using (a) the original Hamiltonian and (b) the WannSymm symmetrized Hamiltonian. Each colors denotes a band structure obtained with one of the symmetrically equivalent high symmetry \textbf{k}-paths. Inset of (a) and (b) are zoomed-in chart showing the details of the combined band structure.} 
  \label{K233-band}
\end{figure}

The Wannier Hamiltonian is generated using 96 states containing Cr-3d and As-4p orbitals in the presence of SOC. Due to the crystal symmetry, there are 12 symmetrically equivalent \textbf{k}-paths for band structure. Using the real-space Wannier Hamiltonian, we calculate the band structure of all the equivalent \textbf{k}-paths and illustrate them in a single plot [Fig. \ref{K233-band} (a)]. In the ideal case, the band structures from these equivalent high symmetry \textbf{k}-paths should be exactly the same. However, the Wannierization procedure has no symmetry constraint and the numerical errors are therefore inevitable. The inset of Fig. \ref{K233-band} (a) shows a zoomed-in part of band structure along K-$\Gamma$, illustrating the poor symmetry of the Hamiltonian without symmetrization. The energy differences are the order of $\sim 3\times10^{-4}$ eV, which is already larger than commonly employed symmetry criterion for double precision calculations (in most cases no larger than $10^{-6}$ eV). With the symmetrized real-space Hamiltonian, we also calculated band structures along all 12 equivalent \textbf{k}-paths, which is illustrated in Fig. \ref{K233-band} (b). Overall, the band structure generated using the symmetrized Hamiltonian resembles the one from original Hamiltonian, and the DFT band structure as well. In addition, the violated symmetry during the wannierization is restored by the symmetrization code, as shown in the inset of Fig. \ref{K233-band} (b). Numerically, the energy differences between the 12 equivalent \textbf{k}-paths reach the order of $10^{-14}$ eV, which is close to the limit of double precision float. 

Using our code, we have also calculated the symmetry characters for the 10 states near the Fermi level (from -0.07 eV to 0.10 eV) at $\Gamma$ point. With the character table given in Ref. \cite{GroupTheoryBradley}, we have also determined the irreducible representation for these states. The results are (1-2) $\Gamma_9$; (3-4) $\Gamma_8$; (5-6) $\Gamma_7$; (7-8) $\Gamma_9$ and (9-10) $\Gamma_8$, the numbers inside the brakets are band indices labeled in the order of increasing band energies, and the two indices within the same brakets are degenerate states. These results are the same as those calculated by Quantum ESPRESSO\cite{Code-QE}. The eigenvalues of symmetry operators (i.e. the parity  at the time-reversal invariant momenta, the mirror eigenvalue and the eigenvalue for rotation symmetry) as reported in our previous studies \cite{Xu-233-cal-topo,W2As3,LI2021243} can also be obtained with this code.
 
\begin{figure}[htbp]
    \hspace*{\fill}
    \begin{subfigure}{0.387\textwidth}
        \begin{subfigure}{1.0\textwidth}
  	        \includegraphics[width=\linewidth]{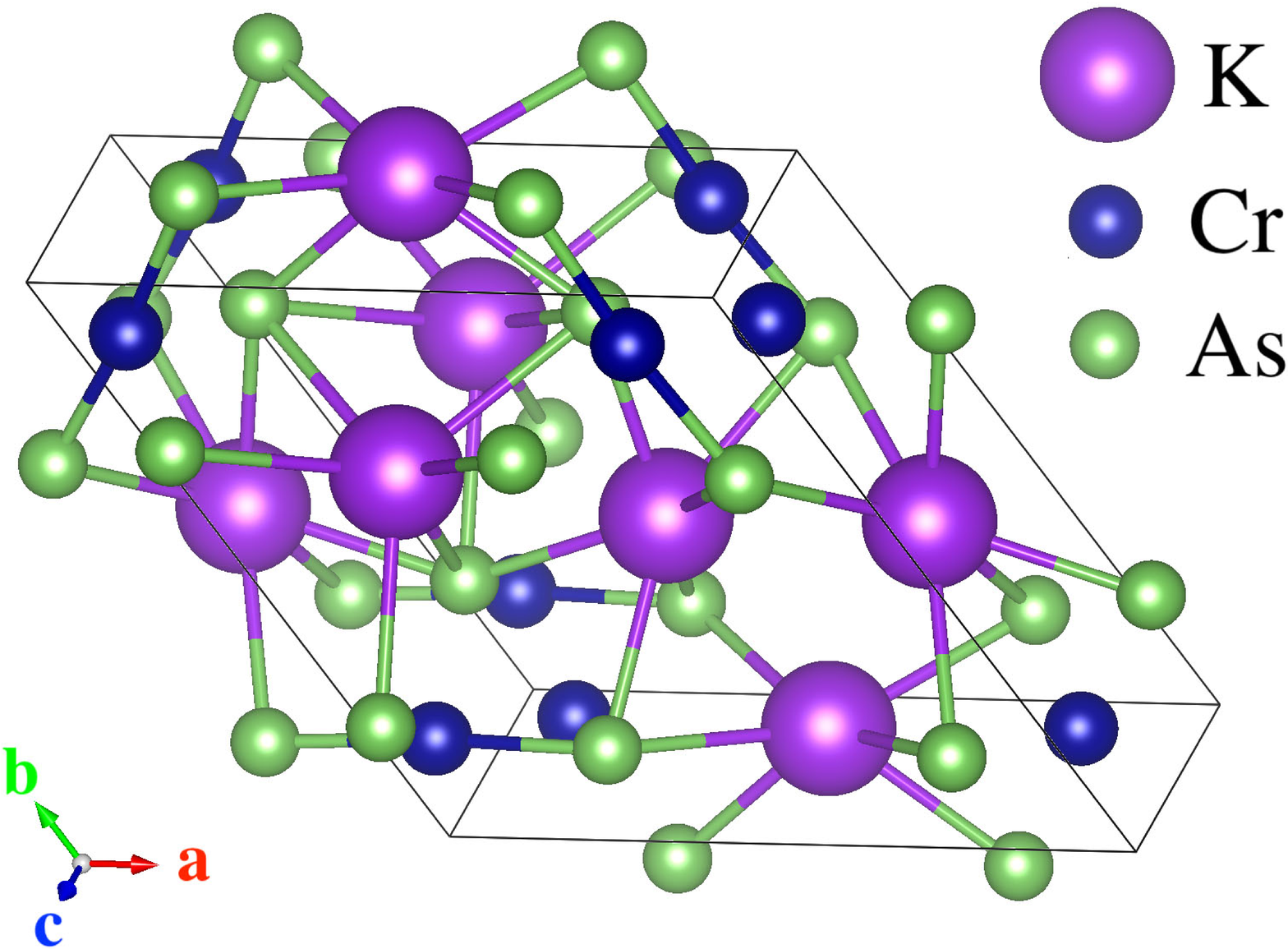}
 	          \caption{} 
        \end{subfigure}
        \begin{subfigure}{1.0\textwidth}
            \label{K233-kpath-label}
            \includegraphics[width=\linewidth]{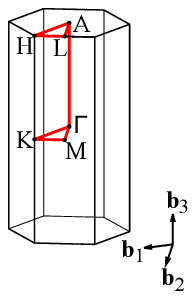}
 	          \caption{}
        \end{subfigure}
    \end{subfigure}
    \begin{subfigure}{0.592\textwidth}
        \begin{subfigure}{1.0\textwidth}
          \includegraphics[width=\linewidth]{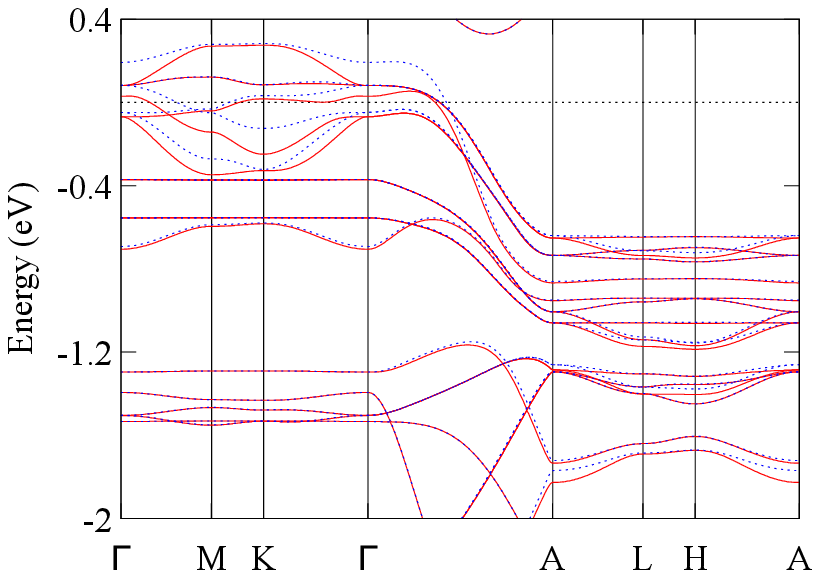}
 	        \caption{} 
        \end{subfigure}
        \begin{subfigure}{1.0\textwidth}
          \includegraphics[width=\linewidth]{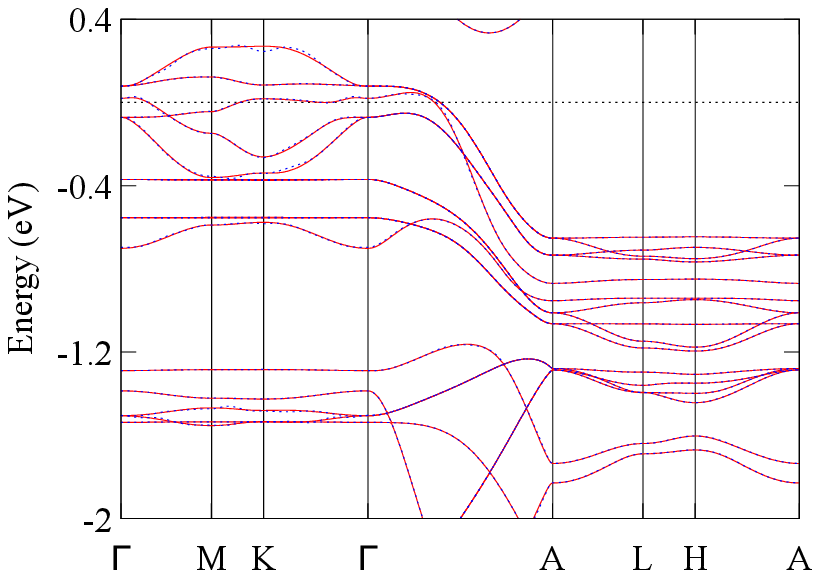}
 	        \caption{} 
        \end{subfigure}
    \end{subfigure}
    \hspace*{\fill}
    \caption{(a) Crystal structure of K$_2$Cr$_3$As$_3$ created with VESTA 3 package \cite{VESTA3}. (b) High-symmetry \textbf{k}-path selected. 
    (c) DFT band structure of K$_2$Cr$_3$As$_3$ (solid red line) and the Wannier interpolated band structure derived from the symmetry-adapted Wannier functions (dotted blue line). 
    (d) DFT band structure of K$_2$Cr$_3$As$_3$ (solid red line) and the band structure reconstructed from the WannSymm symmetrized Hamiltonian (dotted blue line). SOC is not considered in the calculations of (c-d). }
    \label{K233-struct-kpath}
\end{figure}

We want to take a minor detour and discuss about the differences between our method and the symmetry adapted Wannier function method\cite{SymmAdaptedWannFun}. Unlike our method, the symmetry adapted Wannier function method employs the symmetry constraint during the wannierization and minimization procedures. Conceptually, it may be more appealing than postprocess procedures like our method. In practice, however, for some complex systems, the Bl\"{o}ch state may involve too many higher order angular momentum contributions, and the symmetry adapted Wannier function may fail to capture the low-energy excitation. Such an example is illustrated in Fig.\ref{K233-struct-kpath} (c-d), where we compare the interpolated band structures obtained from symmetry-adapted Wannier functions,  symmetrized real-space Hamiltonian and DFT band structures in the absence of SOC. The real-space Wannier Hamiltonian for both cases are obtained  using 48 atomic-like orbitals including Cr-$3d$ and As-$4p$ in non-SOC case. We employ the Wannierization procedure with the same outer energy window and without minimization of Wannier spreading. The symmetrized real-space Hamiltonian using WannSymm correctly reproduces the low-energy excitations as well as the band symmetry at the $\Gamma$ point around the Fermi level, in contrast to the symmetry adapted Wannier function method.

\subsection{Anti-ferromagnet \texorpdfstring{MnF$_2$}{MnF2} and Ferromagnet \texorpdfstring{CrO$_2$}{CrO2}}

\begin{figure}[htbp]
  \begin{subfigure}{0.32\textwidth}
  	\includegraphics[width=\linewidth]{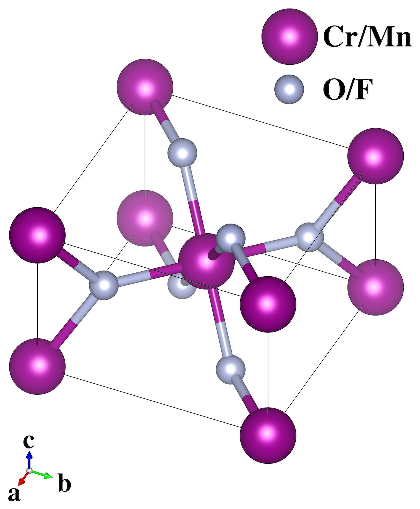}
 	\caption{} 
 	\label{fig:MCStruct-a}
  \end{subfigure}  
  \hspace*{\fill}
  \begin{subfigure}{0.32\textwidth}
  	\includegraphics[width=\linewidth]{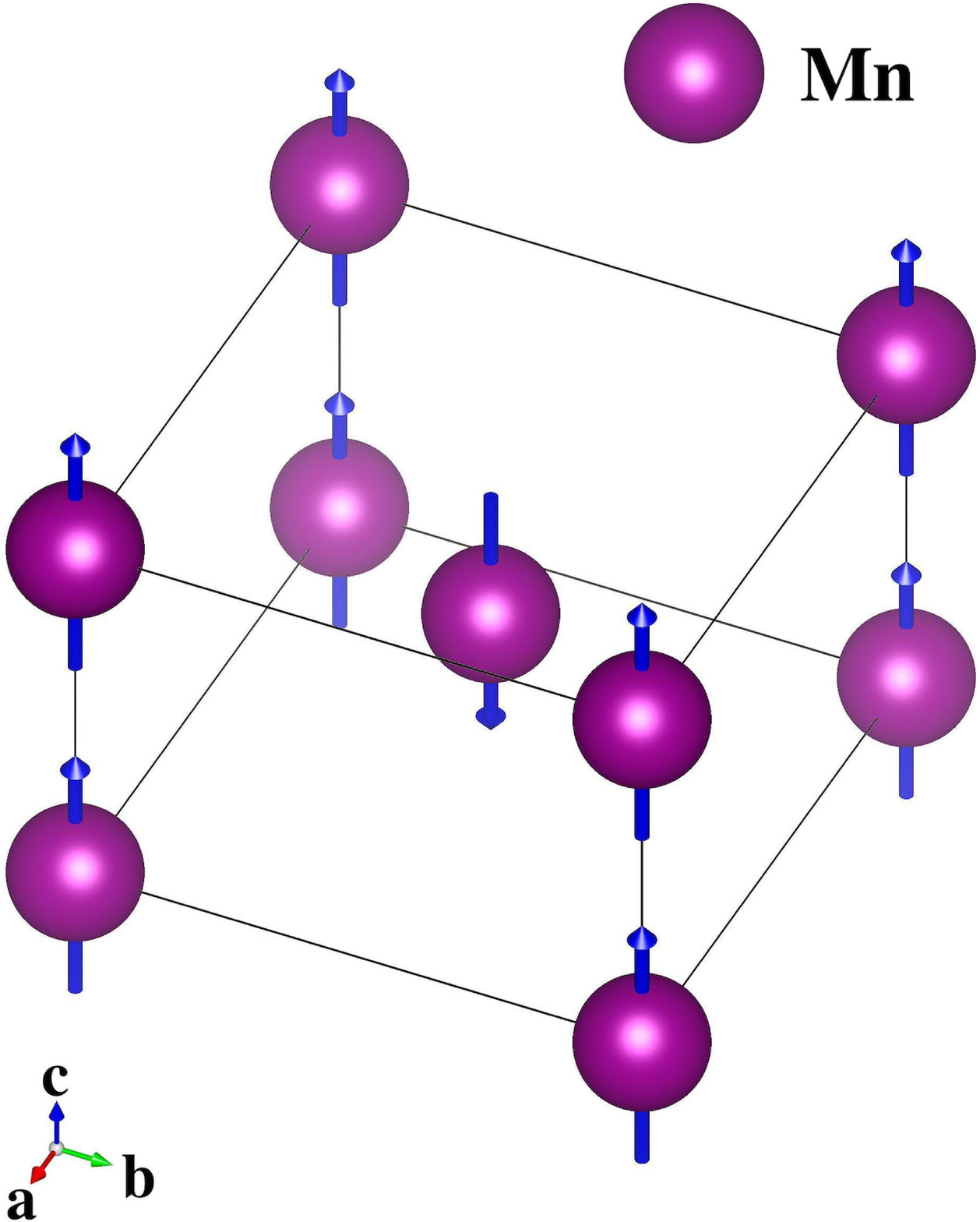}
 	\caption{} 
 	\label{fig:MCStruct-b}
  \end{subfigure}
  \hspace*{\fill} 
  \begin{subfigure}{0.32\textwidth}
  	\includegraphics[width=\linewidth]{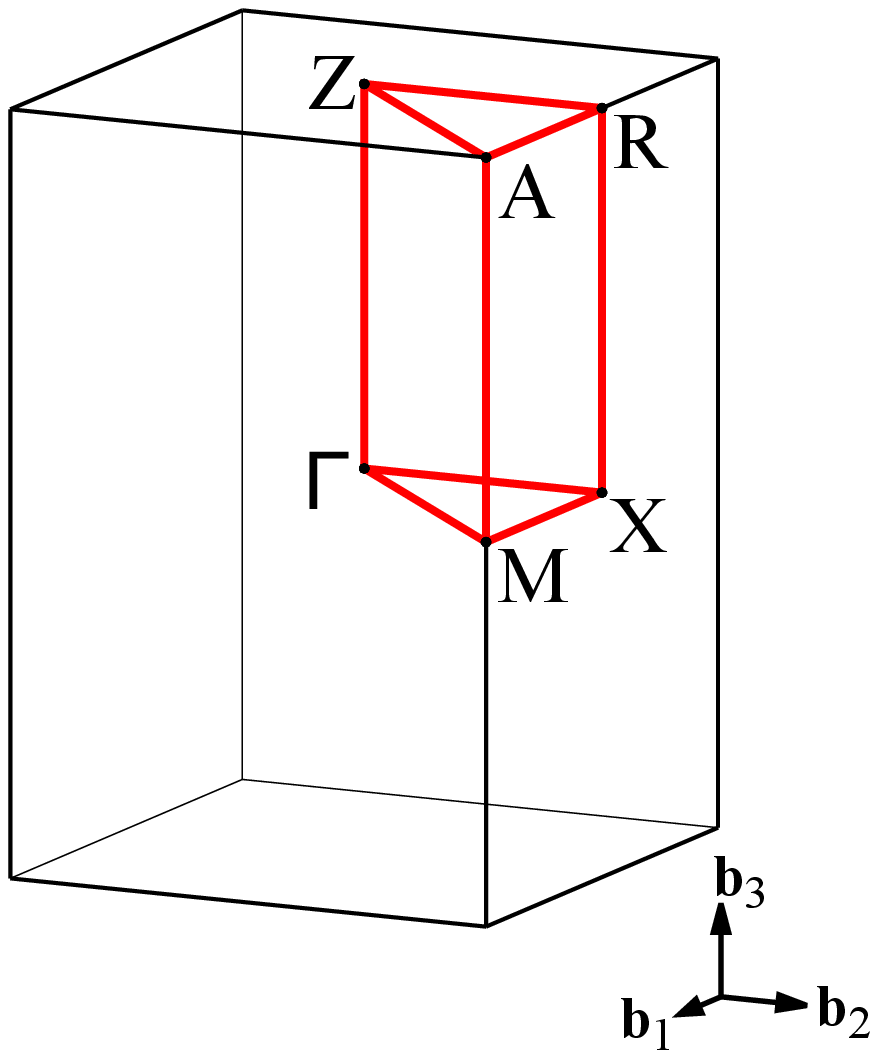}
  	\caption{} 
  	\label{fig:MCStruct-c}
  \end{subfigure}
\caption{(a) Crystal structure of MnF$_2$ and CrO$_2$. (b) Lattice of Mn in MnF$_2$, the arrows show the orientation of magnetic moments on each Mn$^{2+}$. The structures are created with VESTA 3 package \cite{VESTA3}. (c) Illustration of high symmetry \textbf{k}-path of band structure plotting for MnF$_2$ or CrO$_2$. } 
  \label{MnF2CrO2-struct}
\end{figure}

MnF$_2$ is a typical anti-ferromagnetic (AFM) material \cite{MnF2-exp-AFM}. Its space group is $P4_2 /mnm$ (No.\,136), which contains 16 symmetry operations. We show the crystal structure as well as the magnetic pattern in Fig. \ref{MnF2CrO2-struct} (a-b). The original real-space Wannier Hamiltonian $H^{\mathrm{in}}_{\mathrm{MnF_2}}$ is constructed using 48 Wannier states, including the Mn-4s, Mn-3d and F-2p atomic orbitals. 

\begin{figure}[htbp]
  \begin{subfigure}{0.49\textwidth}
  	\includegraphics[width=\linewidth]{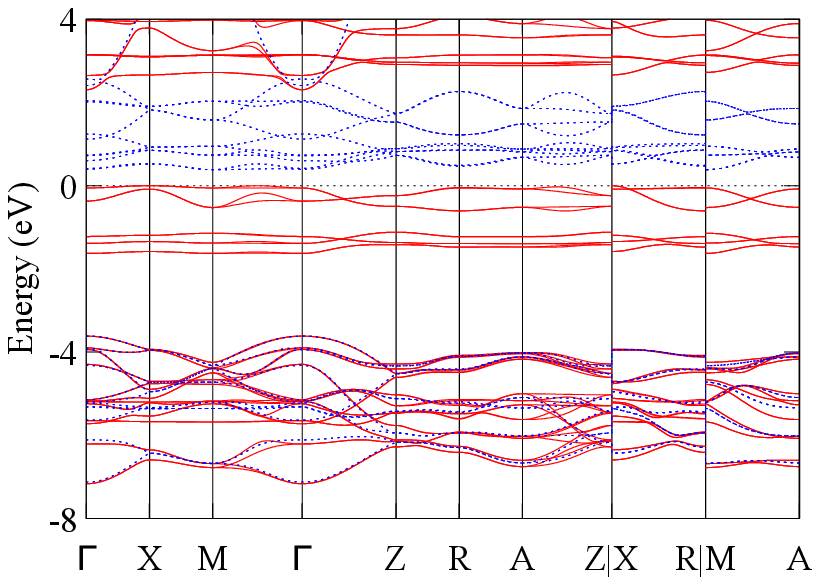}
 	\caption{} \label{fig:MnF2bndb}
  \end{subfigure}
  \hspace*{\fill} 
  \begin{subfigure}{0.49\textwidth}
  	\includegraphics[width=\linewidth]{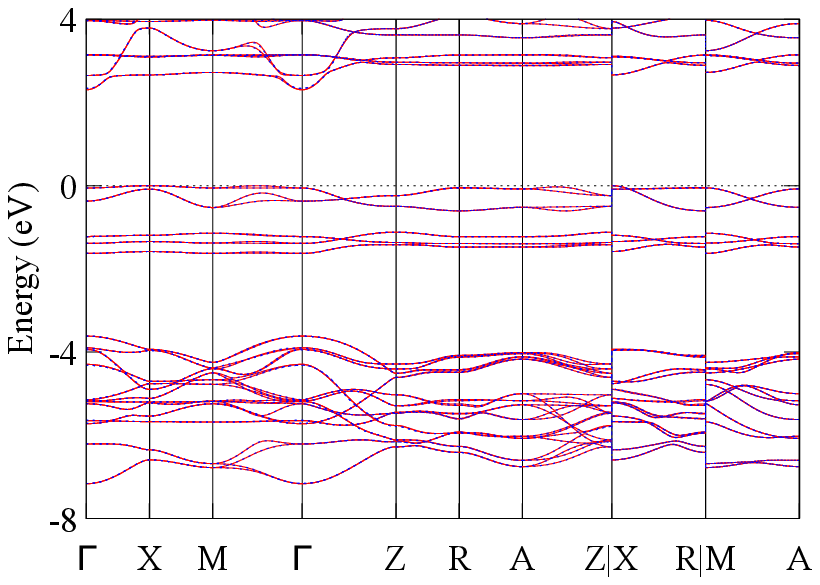}
    \caption{} \label{fig:MnF2bndc}
  \end{subfigure}
\caption{DFT band structure of MnF$_2$ (solid red line) and the band structure reconstructed from the symmetrized Hamiltonian (dotted blue line).
(a) AFM order is not considered in the symmetrization procedure.
(b) AFM order is considered in the symmetrization procedure.
} 
  \label{MnF2-band}
\end{figure}

In the presence of SOC, we show in Fig. \ref{MnF2-band} (a) the original DFT band structure as well as the band structure reconstructed using $H^{\mathrm{nomag}}_{\mathrm{MnF_2}}$, which is derived by symmetrizing $H^{\mathrm{in}}_{\mathrm{MnF_2}}$ without considering the long-range AFM order. With $H^{\mathrm{in}}_{\mathrm{MnF_2}}$, we can calculate the Zeeman splittings for F-2p and the exchange splittings for Mn-d. And the Zeeman splittings for F-2p$_{x(y)}$ and F-2p$_z$ are 0.024 and 0.160 eV, respectively. The former two are equivalent because the magnetic moment is assumed to be parallel to $c$. And the exchange splittings for Mn-d$_{z^{2}}$, Mn-d$_{x^{2}-y^{2}}$, Mn-d$_{zx(y)}$, Mn-d$_{xy}$ are 4.401, 4.486, 4.397 and 4.392 eV, respectively. Again, d$_{zx}$ and d$_{zy}$ are equivalent due to crystal symmetry. The ligand states between [-8, -3] eV are mostly contributed by F-2p orbitals, where only small Zeeman splitting (which is only 3\% as large as the exchange splitting for Mn-d) is present, and are therefore less affected by the ignorance of magnetic order. The Mn-3d states between [-2, 4] eV, however, are apparently inconsistent with the original DFT band structure as the exchange-splitting is (incorrectly) suppressed to 0 eV from $\sim$ 4.39 eV if the global time-reversal symmetry is also considered during the symmetrization process. We then derive the correct magnetic group using the procedure described above, the resulting magnetic group contains 16 symmetry operations, including 1 four-fold rotation symmetry, 4 two-fold rotation symmetry and an inversion symmetry. Using the correct magnetic group, we symmetrize $H^{\mathrm{in}}_{\mathrm{MnF_2}}$ to obtain the real-space Hamiltonian $H^{\mathrm{mag}}_{\mathrm{MnF_2}}$. Fig. \ref{MnF2-band} (b) shows the comparison between the original DFT band structure and the band structure reconstructed using $H^{\mathrm{mag}}_{\mathrm{MnF_2}}$. The good comparison manifests the validity of our method.

\begin{figure}[htbp]
  \begin{subfigure}{0.49\textwidth}
  	\includegraphics[width=\linewidth]{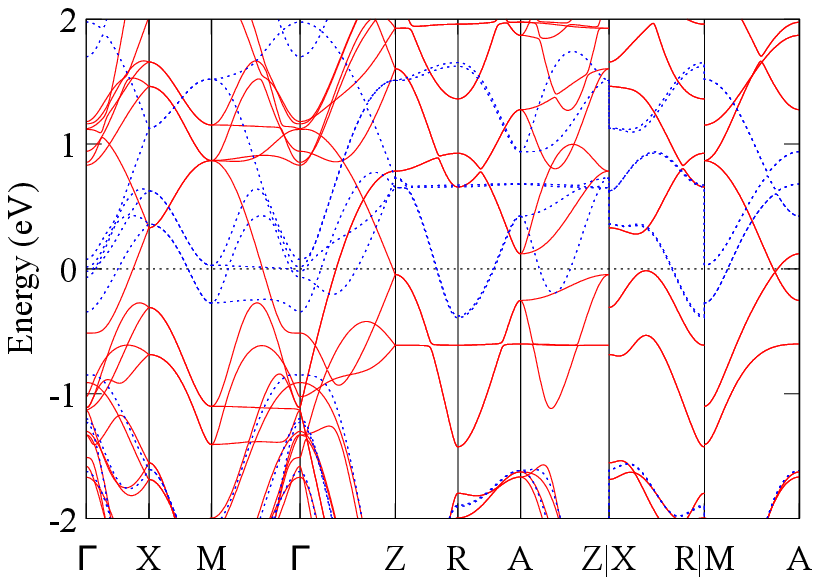}
 	  \caption{} 
 	  \label{CrO2-band:sym1}
  \end{subfigure}
  \hspace*{\fill} 
  \begin{subfigure}{0.49\textwidth}
  	\includegraphics[width=\linewidth]{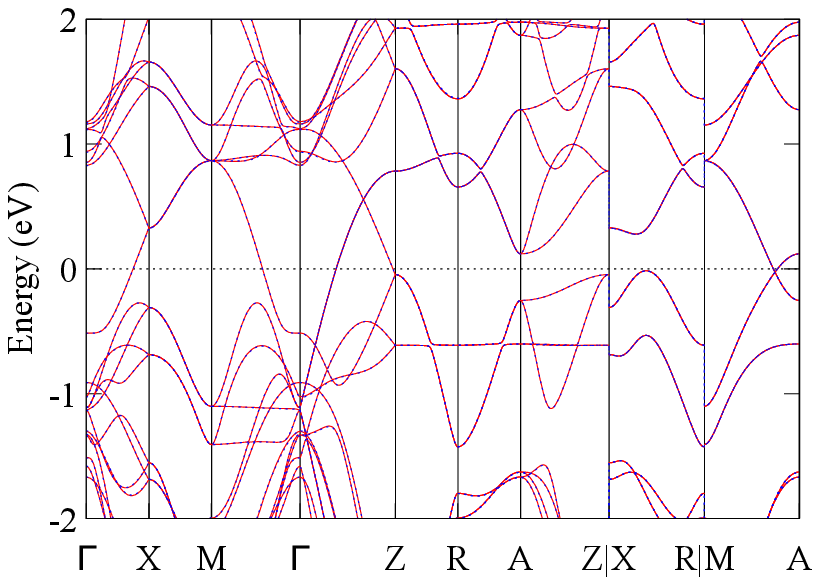}
    \caption{} 
    \label{CrO2-band:sym2}
  \end{subfigure}
\caption{
DFT band structure of CrO$_2$ (solid red line) and band structure reconstructed from the symmetrized Hamiltonian (dotted blue line).
(a) FM order is not considered in the symmetrization procedure.
(b) FM order is considered in the symmetrization procedure.
}

  \label{CrO2-band}
\end{figure}

We illustrate the ferromagnetic (FM) case using CrO$_2$. CrO$_2$ is a typical FM material\cite{CrO2-FM}, it shares the same space group  $P4_2 /mnm$ (No.\,136) with MnF$_2$ and the crystal structure is illustrated in Fig. \ref{MnF2CrO2-struct} (b). For simplicity, we constrain the magnetic moment along $z$-direction in the presence of SOC. Please be advised that the moment direction also affects the determination of magnetic group. In this case, we obtain the original Wannier real-space Hamiltonian ($H^{\mathrm{in}}_{\mathrm{CrO_2}}$) using 44 states including Cr-3d and O-2p orbitals. Similar to the MnF$_2$ case, we also compare the original DFT band structure and band structure reconstructed with symmetrized Hamiltonian with/without considering magnetic order (Fig. \ref{CrO2-band}). In Fig. \ref{CrO2-band}(a), the symmetrization was done without considering FM order, and thus the large exchange splitting ($\sim$ 2 eV for Cr-d) is missing. In contrast, the exchange-splitting is preserved, and the correct band structure is reproduced in Fig. \ref{CrO2-band}(b) once the FM order is taken into consideration.

\subsection{Topological nodal-line semimetal \texorpdfstring{Ce$_3$Pd$_3$Bi$_4$}{Ce3Pd3Bi4}}
\label{SecEgCe343}
\begin{figure}[htbp]
    \begin{subfigure}{0.3\textwidth}
  	\includegraphics[width=\linewidth]{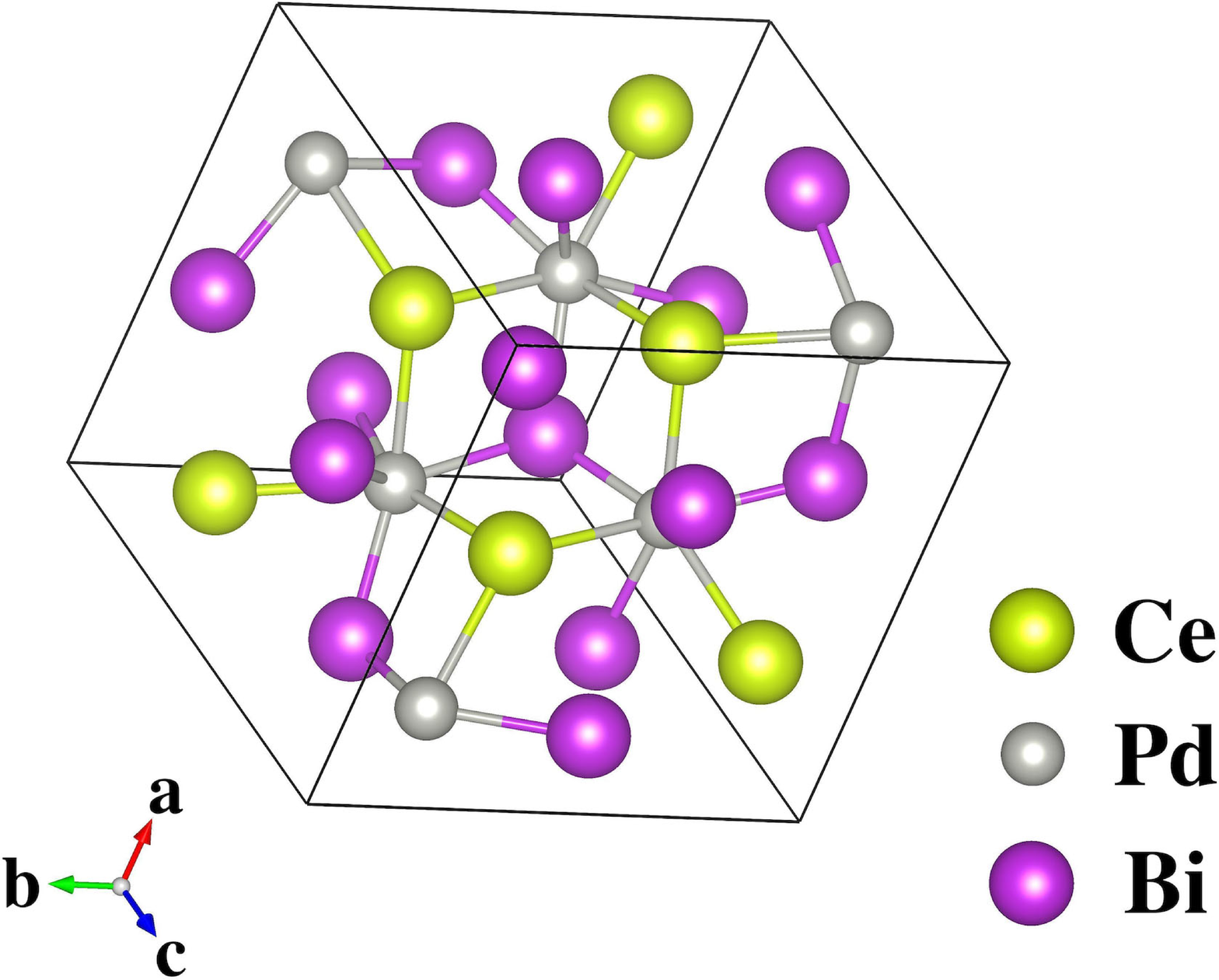}
 	  \caption{} 
  \end{subfigure}
  \hfill
  \begin{subfigure}{0.3\textwidth}
  	\includegraphics[width=\linewidth]{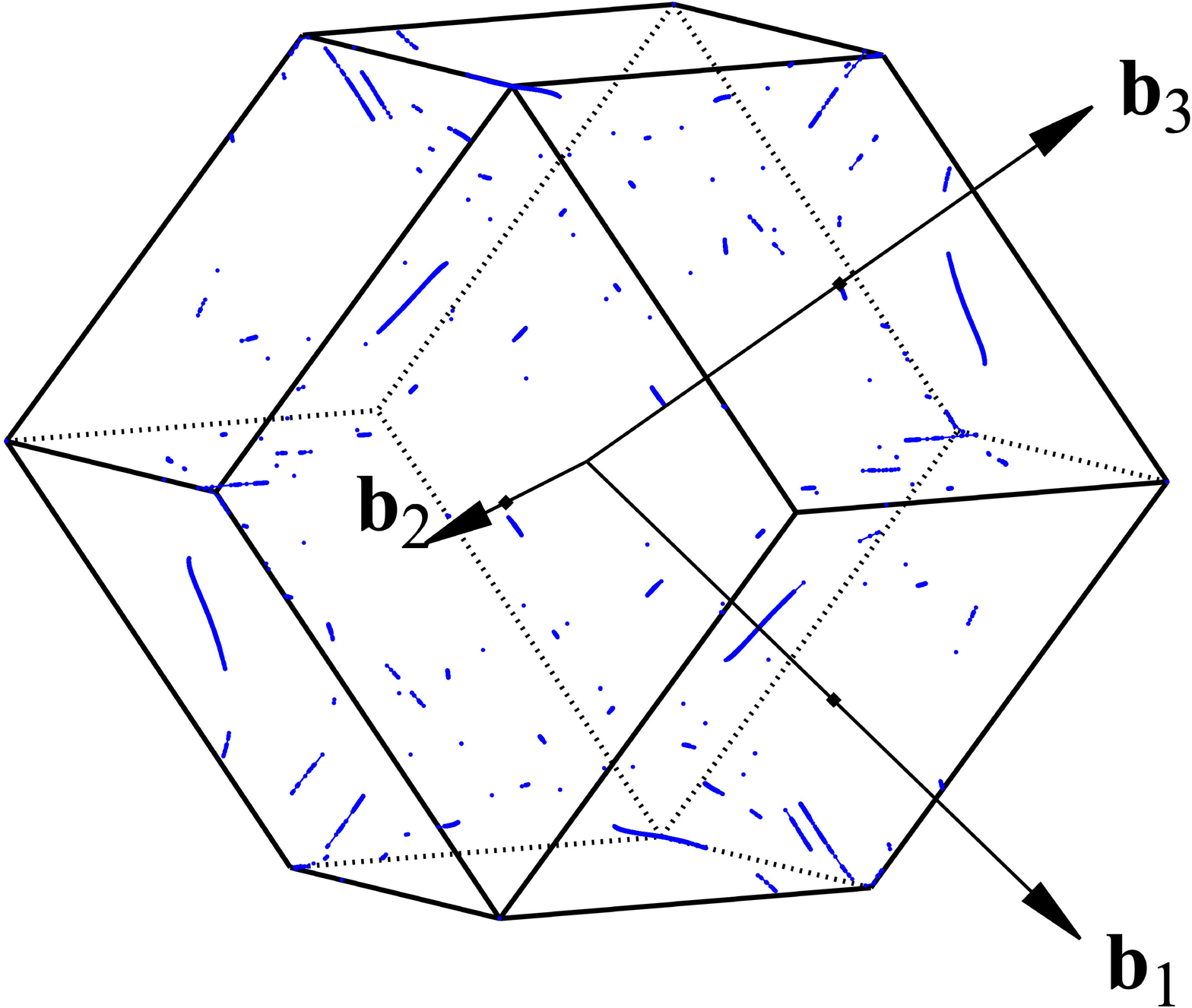}
 	  \caption{} 
  \end{subfigure}
  \hfill
  \begin{subfigure}{0.3\textwidth}
  	\includegraphics[width=\linewidth]{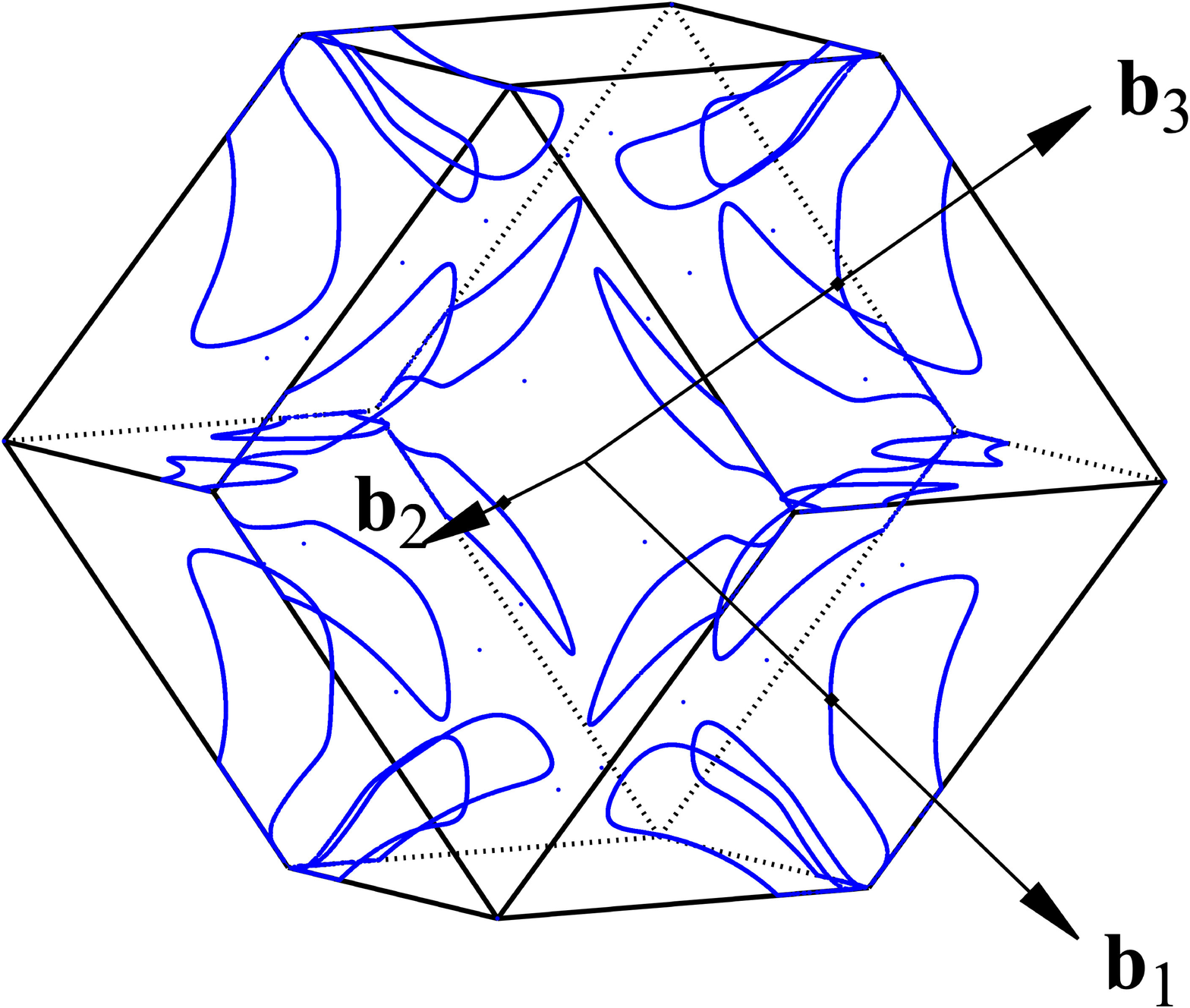}
 	  \caption{} 
  \end{subfigure}
\caption{(a) Primitive unit cell of Ce$_3$Pd$_3$Bi$_4$ plot with VESTA 3 package \cite{VESTA3}. (b-c) The first Brillouin zone showing the nodal points (blue points) and nodal lines (blue lines) of Ce$_3$Pd$_3$Bi$_4$ calculated with (b) the original Hamiltonian and (c) the WannSymm symmetrized Hamiltonian.}
  \label{Ce343-nodal}
\end{figure}

Ce$_3$Pd$_3$Bi$_4$ was recently identified as a topological Kondo nodal-line semimetal\cite{Ce343prl}. It has a body-centered cubic structure with space group $I\bar{4}3d$ (No. 220). Its primitive unit cell, as shown in Fig. \ref{Ce343-nodal} (a), contains 24 symmetries, including 6 gliding mirror symmetries.

The real-space Wannier Hamiltonian ($H^{\mathrm{in}}_{\mathrm{Ce343}}$) is obtained using 108 states including Pd-4d and Bi-3p orbitals. We then employ the WannierTools code\cite{Code-WannierTools} to find the nodal structures between the band 4 and band 5 (band indices follow the Supplemental Material of \cite{Ce343prl}). The resulting nodal lines and nodal points are shown in Fig. \ref{Ce343-nodal} (b). Since the symmetry of $H^{\mathrm{in}}_{\mathrm{Ce343}}$ is poor, it is almost impossible to correctly identify the nodal structures. However, after the symmetrization procedure, the result improves drastically. The resulting nodal structure using the symmetrized real-space Hamiltonian ($H^{\mathrm{sym}}_{\mathrm{Ce343}}$) is shown in Fig. \ref{Ce343-nodal} (c), illustrating the correct nodal rings around the Fermi level. Such a significant change brought by the symmetrization procedure suggests that the topological properties are sensitive to the Hamiltonian symmetry, and thus a Hamiltonian with correct symmetry is crucial to the topological classification. 

\section{Conclusion}
WannSymm is a user friendly, fast, MPI-parallelized open source C-based code for symmetry analysis based on real-space Hamiltonian constructed using atomic-like orbitals. We implement not only for the nonmagnetic cases, but also for the magnetic cases without and with SOC. We demonstrate the validity of our method by using typical materials, e.g. K$_2$Cr$_3$As$_3$, MnF$_2$, CrO$_2$, and Ce$_3$Pd$_3$Bi$_4$. After symmetrization, the crystal symmetries ruined during the Wannierise procedure are restored, and thus the improved numerical results can be obtained. Eigenvalues and characters of an arbitrary symmetry operator (symmorphic or nonsymmorphic) can also be obtained using this code. Since our code directly handles real-space Hamiltonian output from Wannier90 code, it can easily be used in combination with any code that has an interface to Wannier90, including VASP, Quantum ESPRESSO and Wien2K. We emphasize here that our current method can be applied to linear combinations of atomic orbitals (e.g., sp$^2$-like, sp$^3$-like orbitals, etc) with small modifications, but cannot be applied to general Wannier orbitals with arbitrary angular dependence (e.g. summation of Bl\"{o}ch waves). Finally, the method does not depend on the details of Wannier orbital generation except for atomic-like orbital assumption, and therefore can be applied to any real-space Hamiltonian as long as it is defined on a basis set which has atomic-like angular dependence.

\section*{Acknowledgements}
The authors would like to thank Jian-Xin Zhu for stimulating discussions about the magnetic cases and testing the code. The project was support by NSFC (No.s 11874137 and 12074333),  National Basic Research Program of China (No.s 2016YFA0300402 and 2014CB648400), and the Key R\&D Program of Zhejiang Province China (No. 2021C01002). The calculations were performed at the High Performance Computing Cluster of Center of Correlated Matter at Zhejiang University and Beijing SuperComputing Center.


\bibliography{wannsymm}

\end{document}